\documentclass[
aps,
amsmath,
amssymb,
prapplied,
twocolumn,
groupedaddress,
10pt,
]{revtex4-2}

\usepackage{graphicx}
\usepackage{dcolumn}
\usepackage{bm}
\usepackage{float}
\usepackage{hyperref}
\usepackage[mathlines]{lineno}
\hypersetup{colorlinks=true, linkcolor=blue, anchorcolor=blue, citecolor=blue}
\DeclareUnicodeCharacter{0301}{\'{e}}

\begin{document}
	
	
	\title{Polarization-entangled quantum frequency comb \\
		from a silicon nitride microring resonator}
	
	\author{Wenjun Wen,$^{1}$ Wenhan Yan,$^{1}$ Chi Lu,$^{1}$ Liangliang Lu,$^{2,1}$ Xiaoyu Wu,$^{1}$ Yanqing Lu,$^{1}$ Shining Zhu,$^{1}$ and Xiao-song Ma$^{1,3,4,\ast}$}
	
	\affiliation{
		$^1$National Laboratory of Solid-state Microstructures, School of Physics, College of Engineering and Applied Sciences, Collaborative Innovation Center of Advanced Microstructures, Nanjing University, Nanjing 210093, China\\
		$^2$Key Laboratory of Optoelectronic Technology of Jiangsu Province, School of Physical Science and Technology, Nanjing Normal University, Nanjing 210023, China\\
		$^3$Synergetic Innovation Center of Quantum Information and Quantum Physics, University of Science and Technology of China, Hefei, Anhui 230026, China\\
		$^4$Hefei National Laboratory, Hefei 230088, China\\
		$^{\ast}$e-mails: Xiaosong.Ma@nju.edu.cn}
	
	\date{\today}
	
	\begin{abstract}
		Integrated micro-resonator facilitates the realization of quantum frequency comb (QFC), which provides a large number of discrete frequency modes with broadband spectral range and narrow linewidth. However, all previous demonstrations have focused on the generation of energy-time or time-bin entangled photons from QFC. Realizing polarization-entangled quantum frequency comb, which is the important resource for fundamental study of quantum mechanics and quantum information applications, remains challenging. Here, we demonstrate, for the first time, a broadband polarization-entangled quantum frequency comb by combining an integrated silicon nitride micro-resonator with a Sagnac interferometer. With a free spectral range of about $99 ~{\rm GHz}$ and a narrow linewidth of about $190 ~{\rm MHz}$, our source provides 22 polarization entangled photons pairs with frequency covering the whole telecom C-band. The entanglement fidelities for all 22 pairs are above $81\%$, including 17 pairs with fidelities higher than $90\%$. Our demonstration paves the way for employing the polarization-entangled quantum frequency comb in quantum network using CMOS technology as well as standard dense wavelength division multiplexing technology.
	\end{abstract}
	
	\maketitle
	
	\section{Introduction}
	
	Entangled photons play a crucial role in the fundamentals of physics \cite{Clauser1978bell, Aspect1982experimental, Zeilinger1990bell, Hensen2015loophole, Giustina2015significant, Shalm2015strong, Rosenfeld2017event} and have practical applications in quantum information science, such as quantum communication and quantum cryptography and so on \cite{Ursin2007entanglement, Fedrizzi2009high, Yin2017satellite, Yin2017prl, Ren2017ground}. In particular, entangled photons are used to share secret cryptographic keys between distant parties with quantum key distribution (QKD) protocols. One can combine various degrees of freedoms (DOFs) of photons, which has great potential to expand towards higher dimensionality \cite{Xie2015harnessing, Joshi2018frequency}. Quantum networks \cite{Wengerowsky2018entanglement, Joshi2020, liu202240, Wen2022} (QNs) with high efficiency and scalability in real-world scenarios are expected to be realized with entangled photons.
	
	Recently, with advancements of integrated optics technologies, such as flexible dispersion engineering, high-quality ($Q$) factor, small-mode volume and compatible with the CMOS fabrication process, optical microresonators \cite{Chang2022integrated} become robust, scalable and reconfigurable devices for quantum optical applications. Based on spontaneous four-wave mixing (SFWM) using $\chi^{3}$ nonlinearity \cite{Li2005optical, Helt2010spontaneous}, microresonators have been used to generate energy-time \cite{Wen2022, Mazeas2016high, Jaramillo2017persistent, Fan2023Multi}, time-bin \cite{Samara2019amplified, Samara2021Entanglement} and frequency-bin \cite{Chen2011Frequency, Imany2018} entangled photons. Meanwhile, due to the inherent multiresonant mode properties of microresonators, integrated cavity-enhanced devices can directly generate quantum frequency combs (QFCs) without the need for subsequent periodic filtering \cite{Reimer2014Integrated, Reimer2016Generation, Kues2019quantum}. In a micro-resonator, the SFWM efficiency for generating photon pairs is proportional to $Q^{3}$ \cite{Guo2018Generation}. However, for QKD protocol using phase or time-bin encoding, a higher $Q$ factor leads to a long photon coherence time and requires a large difference in interferometer arm length, which places stringent requirements on maintaining interferometric phase stability \cite{Lago2021telecom}.
	
	Polarization entanglement is easy to control and analyze, and has mature automatic polarization feedback system, which is a common choice for QKD \cite{Poppe2004Practical, Tang2014Experimental}. One way to obtain polarization entanglement is the Sagnac interferometer \cite{Kim2006Phase, Fedrizzi2007, Wengerowsky2019Entanglement, Liu2020Drone, Yin2020entanglement, Tim2020Long, Neumann2022experimental, Yamazaki2022Massive}, in which two pump beams, one clockwise and one counterclockwise, travel along the same pathway without active phase stabilization between the different directions. In this work, we use a micro-resonator placed in a Sagnac loop to realize a polarization-entangled QFC.
	
	We demonstrate the generation of broadband polarization-entangled QFC in a Sagnac configuration using SFWM on a dispersion-engineered silicon-nitride microring resonator (MRR). With a total of $22$ parallel pairs of frequency channels (limited by the bandwidth of our programmable filter system), a free spectral range of approximately $ 99 ~{\rm GHz}$ and a linewidth of approximately $ 190 ~{\rm MHz}$, our source generates photon pairs in a spectral range of approximately $ 38 ~{\rm nm}$, covering the whole telecom C-band. Long-term operation of our source is ensured by temperature control and feedback control of the high-$Q$ MRR. By analyzing the photonic noise inside and outside the ring, we obtain an effective photon-pair generation bandwidth of approximately $ 3 ~{\rm THz}$. We show that, our source maintains entanglement fidelity above $81\%$ for all $22$ pairs of frequency channels, including 17 pairs with fidelities above $90\%$. Our demonstration suggests that a multimode polarization-entangled biphoton QFC generated by integrated MRR in a Sagnac interferometer can contribute to a promising platform for the realization of multiuser QNs using a commercially available dense wavelength division-multiplexing (DWDM) fiber optical communication network and interfacing solid-state quantum memories at telecommunication wavelengths \cite{jiang2022quantum}.
	
	\begin{figure*}[ht]
		\centering
		\includegraphics[width=0.8\textwidth]{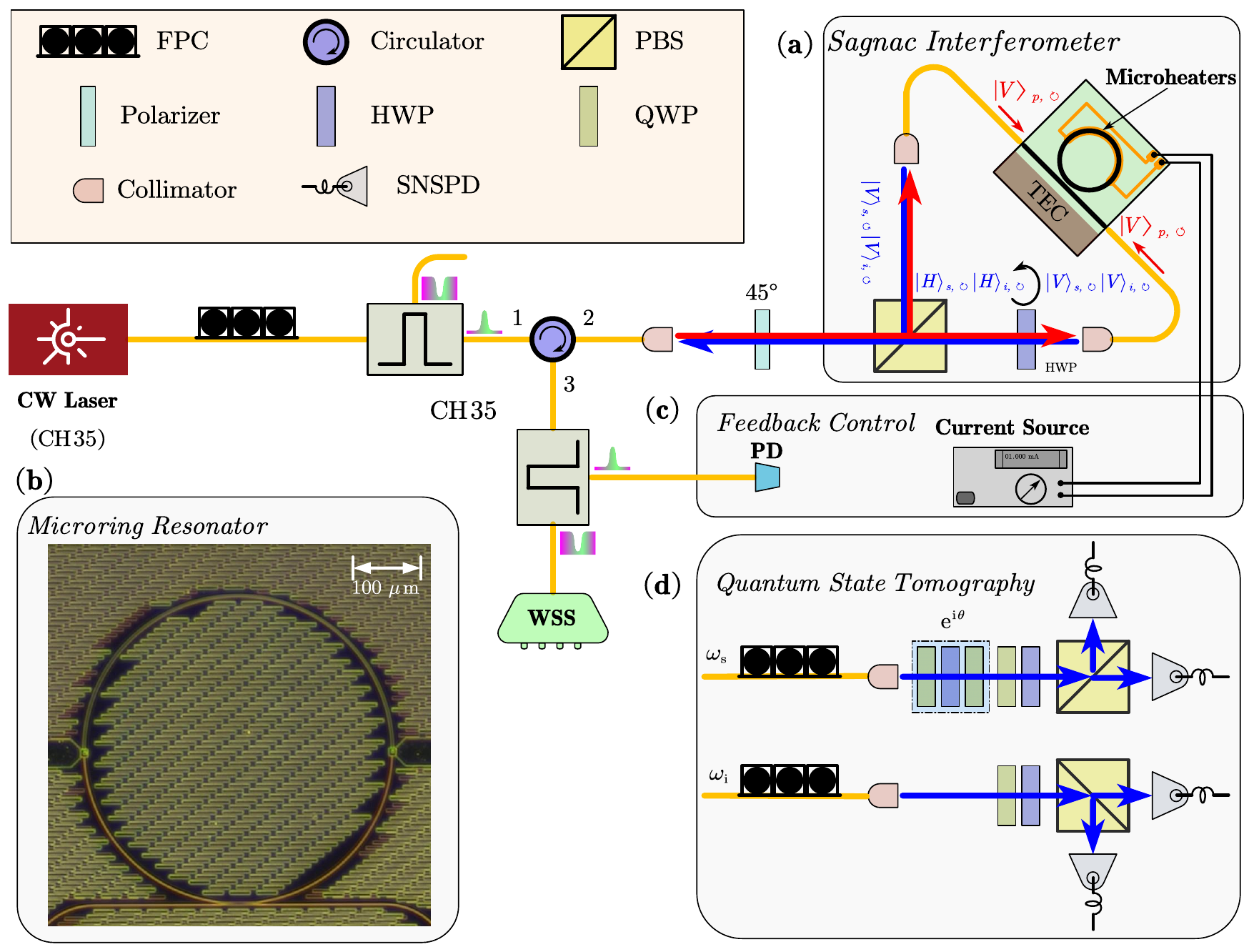}
		\caption{\label{Fig1}Experimental setup for the generation and characterization of the polarization-entangled QFC. (a) Polarization-entangled QFC is generated in a Sagnac configuration, with an integrated silicon-nitride microring resonator at the center of the loop. Wave plates are simplified here. The chip is mounted on a copper plate whose temperature is controlled with high precision ($<0.001~^{\circ} \mathrm{C}$) by a thermoelectric cooler and a negative temperature coefficient (NTC) thermistor. (b) Optical microscopy image of the MRR and the integrated microheater. (c) The cavity resonance is stabilized by using the integrated microheater and feedback control module. (d) Quantum state tomography is performed by the polarization analysis modules. CW, continuous wave; FPC, fiber polarization controller; CIR, circulator; PBS, polarization beam splitter; HWP, half-wave plate; QWP, quarter-wave plate; TEC, thermoelectric cooler; PD, power detector; WSS, wavelength selective switch; SNSPD, superconducting nanowire single-photon detector.}
	\end{figure*}
	
	\section{Experimental Setup}
	
	The schematic diagram of our experimental setup is shown in FIG.\ref{Fig1}. Our quantum light source is based on an integrated Si$_{3}$N$_{4}$ MRR (Ligentec AN800) placed at the center of a Sagnac loop constructed with fiber and bulk optical elements. A cw pump laser (Santec TSL570) at $193.5 ~{\rm THz}$ ($1549.3150 ~{\rm nm}$) is polarized and then coupled into a free-space Sagnac interferometer. A three-port circulator (CIR) placed in front of the loop is used to transfer the pump from port 2 into the loop and to output the residual pump and generated photons through port 3. Multiple DWDMs are cascaded to form the pre- and postfilters used to remove the amplified spontaneous emission of the pump and the residual pump before and after the CIR, respectively.
	
	As shown in FIG.\ref{Fig1}(a), the Sagnac configuration consists of a polarization beam splitter (PBS), an optically packaged SiN MRR chip and several wave plates. The pump incident on the PBS is diagonally polarized and thus split into two orthogonally polarized components ${\left| H \right> _{p,\circlearrowleft}}$ and ${\left| V \right> _{p,\circlearrowright}}$, with horizontally polarized photons propagating counterclockwise and vertically polarized pump photons propagating clockwise. The MRR is designed to be pumped bidirectionally with the vertical polarization mode ${\left| V \right>}$. To achieve this, wave plates are used to adjust the polarization of these two components before they are coupled into the chip. The clockwise component remains vertically polarized ${\left| V \right> _{p,\circlearrowright}}$, while the counterclockwise component is rotated by $90^{\circ}$ (${\left| H \right> _{p,\circlearrowleft}} \rightarrow {\left| V \right> _{p,\circlearrowleft}}$).
	
	Within the MRR, vertically polarized pump photons are converted into vertically polarized photon pairs through SFWM, i.e., ${\left| V \right> _{p}\left| V \right> _{p}} \rightarrow {\left| V \right> _{s}\left| V \right> _{i}}$. The clockwise propagating photon pair ${\left| V \right> _{s,\circlearrowleft} \left| V \right> _{i,\circlearrowleft}}$ are then rotated by $90^{\circ}$ to ${\left| H \right> _{s,\circlearrowleft} \left| H \right> _{i,\circlearrowleft}}$, while the counterclockwise propagating photon pairs ${\left| V \right> _{s,\circlearrowright} \left| V \right> _{i,\circlearrowright}}$ remain unrotated. Here, the subscripts $s$ and $i$ denote signal and idler photons, respectively. Photon pairs generated in clockwise and counterclockwise directions are combined by PBS and exit from the Sagnac interferometer into the input spatial mode. Meanwhile, QFC formed by frequency-entangled photon pairs, i.e. a biphoton QFC, is generated inside the MRR. In the Sagnac interferometer, clockwise and counterclockwise QFCs are generated by the same MRR, ensuring perfect spectral overlap between them. Finally, we obtain a polarization-entangled QFC with the following states \cite{Yamazaki2022Massive}:
	\begin{widetext}
		\begin{equation}
			|\psi\rangle=\frac{1}{\sqrt{2}}\left(|H\rangle_s|H\rangle_i+e^{i \theta}|V\rangle_s|V\rangle_i\right) \otimes \frac{1}{\sqrt{M}}\left(\sum_{m=1}^M|\omega\rangle_{s, m}|\omega\rangle_{i, m}\right),
		\end{equation}
	\end{widetext}
	where $\theta$ is an unknown but constant phase, and we compensate for this undetermined phase using a sandwiched QWP-HWP-QWP setup, i.e. orienting all QWPs at $45^{\circ}$ and rotating the HWP \cite{Lingaraju2021Adaptive, Muneer2021Reconfigurable}. And ${\left| \omega \right> _{s(i),m}}$ represents the state of the signal (idler) photon at frequency pair index $m$ ($ m=1,2,\cdots,M $).
	
	The photons leaving port 3 of the CIR contain both the two generated photons and the residual pump, which are separated by the postfilter with a total pump rejection of approximately $ 115 ~{\rm dB}$. The filtered pump is detected by a power detector (PD), which monitors the resonance information of the MRR with respect to the pump. The PD signal is used, in conjunction with the integrated microheater \cite{Mahmudlu2023Fully} electrically driven by a current source (NI 9266), to actively stabilize the resonance of the MRR via thermoelectric tuning.
	
	A C-band wavelength selective switch (WSS) \cite{Muneer2021Reconfigurable} is used to demultiplex the polarization entangled QFC and provide additional pump rejection. To characterize the photon-pair generation, the WSS is configured as a dual bandpass filter to demultiplex signal and idler photons, and send each directly to a detector. The separated signal and idler photons are then sent to two polarization analysis modules (PAMs) for quantum state tomography (QST) \cite{Daniel2001Measurement}. It can also be programmed as a single-channel tunable bandpass filter, which is used to obtain the single-photon spectrum and to analyze the sources of noise.
	
	\begin{figure*}[ht]
		\centering
		\includegraphics[width=0.8\textwidth]{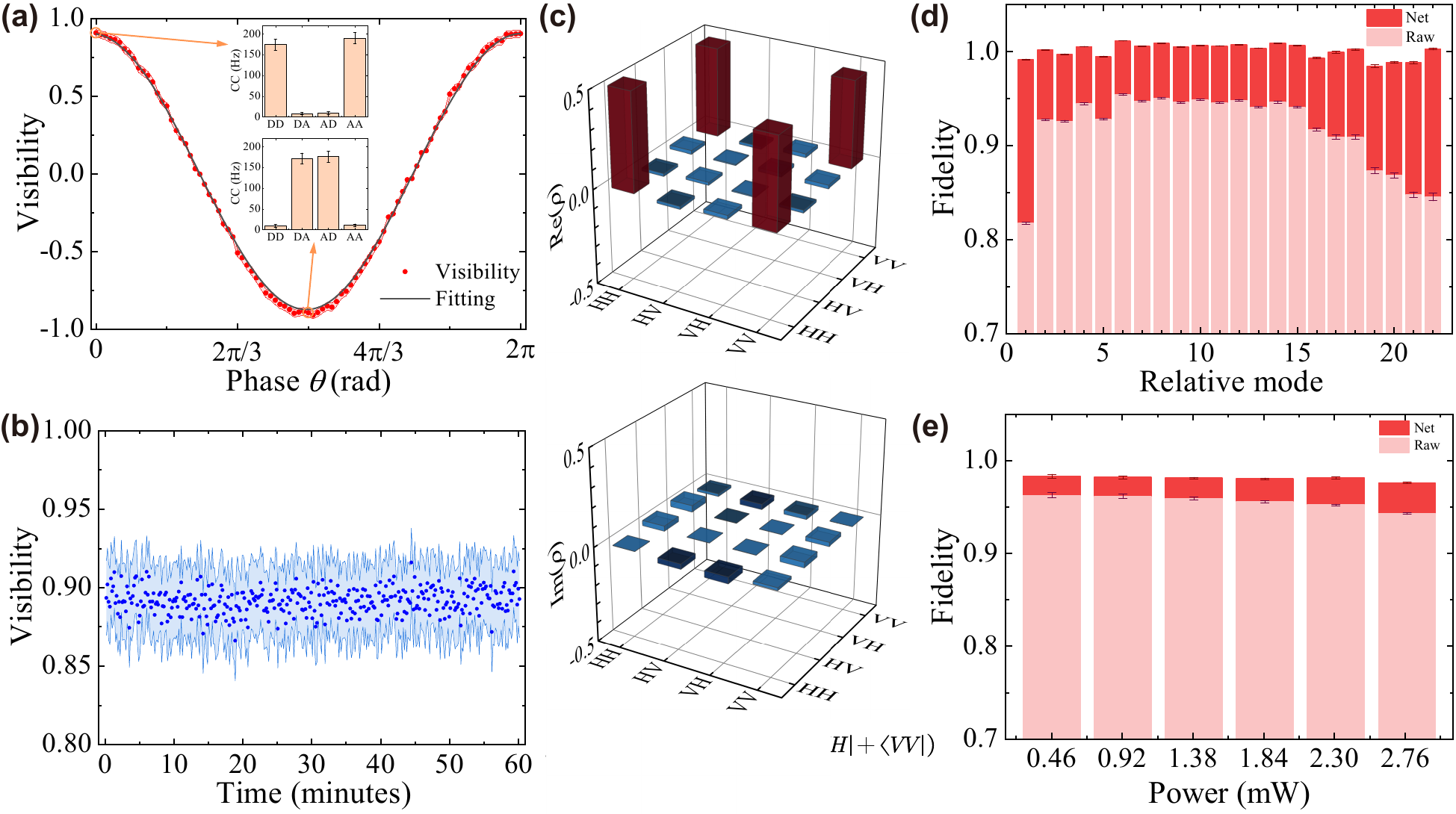}
		\caption{\label{Fig2}(a) Visibility in the ${\left| D \right> } / {\left| A \right> }$ basis as a function of the phase $\theta$. The insets show the two-photon coincidence counts for $\theta=0$ and $\theta=\pi$ in the DA basis, respectively. (b) Visibility stability at $\theta=0$. The shading in (a) and (b) represents the uncertainty in visibilities derived from Poisson statistics. (c) The real and imaginary part of the reconstructed density matrices $\rho$ for the eighth frequency pair ($194.292 - 192.708 ~{\rm THz}$). (d) Fidelities of the polarization entangled states for all $22$ pairs. (e) Fidelity as a function of the pump power for the eighth frequency pair. The uncertainties of the fidelities in (d) and (e) are derived from Poissonian statistics.}
	\end{figure*}
	
	\section{MRR-based broadband polarization entanglement source}
	
	The radius of our MRR is about $230 ~{\rm \mu m}$, while its waveguide cross section is about $0.8 \times 1.6 ~{\rm \mu m ^2}$. Inverted tapers are designed to couple light in and out of the chip, and are end coupled to ultrahigh numerical aperture (UHNA7) fiber arrays, with UV glue curing in between. The measured fiber-chip-fiber insertion loss is about $6 ~{\rm dB}$, including propagation loss. The MRR under test has an average FWHM of $190.41 ~{\rm MHz}$ and an average $Q$ factor of about $1.03\times 10^{6}$, with a free spectral range (FSR) of $99.03 ~{\rm GHz}$ (refer to Appendix A for details).
	
	When using microresonators with high-quality factors and small mode volumes, thermal-induced phenomena such as hysteretic wavelength response \cite{Carmon2004Dynamical} shows up. The hysteresis behavior broadens the cavity lineshape, but at the same time makes it easier for the microresonator to be out of resonance. As shown in FIG.\ref{Fig1}(c), we developed the feedback control based on the transmitted pump power and the integrated microheater, combined with the high-precision temperature control of the chip, to ensure the long-term stable operation of the source. Refer to Appendices B-D for more information.
	
	The frequency correlated photon pairs are spatially separated by a programmable WSS, and each sent to a PAM consisting of a fiber polarization controller (FPC), a quarter-wave plate, a half-wave plate, a polarization beam-splitter cube and two SNSPDs. To map the ${\left| H \right> } / {\left| V \right> }$ basis of the source to the ${\left| H \right> } / {\left| V \right> }$ basis of the PAM, the FPC is used to compensate for the polarization rotation that occurs during transmission between the source and the QWP due to the random birefringence effect of the single-mode fiber. The sandwiched QWP-HWP-QWP setup placed in the signal path is then used to compensate for the remaining phase between $H$ and $V$. We set both QWP to $45^{\circ}$ with respect to horizontal, and then rotate the HWP in the diagonal-antidiagonal (${\left| D \right> } / {\left| A \right> }$) basis, and $\left| D \right> =\left( \left| H \right> +\left| V \right> \right) /\sqrt{2}$, $\left| A \right> =\left( \left| H \right> -\left| V \right> \right) /\sqrt{2}$. Taking the eighth frequency pair ($194.292 - 192.708 ~{\rm THz}$) as an example, the interference curve of visibility as a function of phase $\theta$ is shown in FIG.\ref{Fig2}(a). The visibility is obtained from the coincidence counts measured by the four photon detectors, expressed as
	\begin{equation}
		V=\frac{\left( CC\left( DD \right) +CC\left( AA \right) \right) -\left( CC\left( DA \right) +CC\left( AD \right) \right)}{CC\left( DD \right) +CC\left( AA \right) +CC\left( DA \right) +CC\left( AD \right)},
	\end{equation}
	where $CC(DD)$, $CC(DA)$, $CC(AD)$, and $CC(AA)$ is the coincidence counts in the ${\left| D \right> } / {\left| D \right> }$, ${\left| D \right> } / {\left| A \right> }$, ${\left| A \right> } / {\left| D \right> }$, and ${\left| A \right> } / {\left| A \right> }$ bases respectively. In our experiment, we optimize the angle of the HWP so that the phase between $H$ and $V$ is $\theta=0$ and the ideal quantum state satisfies $\left| \psi \right> \propto \left( \left| HH \right> +\left| VV \right> \right)$. With the temperature and the microheater-based feedback controls, this polarization state can be stably maintained over long time, as shown in FIG.\ref{Fig2}(b).
	
	We then perform QST of the polarization DOF for all frequency correlated pairs in the bases of  ${\left| H \right> } / {\left| V \right> }$,  ${\left| D \right> } / {\left| A \right> }$ and  ${\left| R \right> } / {\left| L \right> }$, where $\left| R \right> =\left( \left| H \right> +i\left| V \right> \right) /\sqrt{2}$, $\left| L \right> =\left( \left| H \right> -i\left| V \right> \right) /\sqrt{2}$. Note that each PMA is equipped with two detectors, and the mutually orthogonal bases can be measured simultaneously. For example, all four basis states, $\left| HH \right>$, $\left| HV \right>$, $\left| VH \right>$, and $\left| VV \right>$, can be obtained in a single measurement. The fidelity to the $\left| \varPsi ^+ \right> =\frac{1}{\sqrt{2}}\left( \left| HH \right> +\left| VV \right> \right)$ Bell state is estimated by coincidence measurements in nine different measurement settings. FIG.\ref{Fig2}(c) shows an example of the reconstructed density matrices for the eighth frequency pair ($194.292 - 192.708 ~{\rm THz}$). For all $22$ frequency pairs, FIG.\ref{Fig2}(d) shows that the observed fidelities are above $81.5\%$ over a spectral range of $4.356 ~{\rm THz}$, among which there are 17 pairs of frequency channels with fidelities higher than $90\%$. The fidelity of the first frequency pair is lower than the others, probably because of its proximity to the pump. Fidelity decreases as the channel is farther away from the pump (still above $84.5\%$). This may be due to mechanisms such as limited SFWM bandwidth, unavoidable noise, etc., which we analyze further in Section IV. After subtracting the accidental coincidences, all net fidelities are above $94\%$. As shown in FIG.\ref{Fig2}(e), when the pump power increases, the fidelity decreases due to multiphoton emission events in SFWM \cite{Li2017On}.
	
	\section{Photonic noise analysis and QFC source characterization}
	
	When pumped by the laser, photons can be generated not only by the SFWM process, i.e. generating correlated photon pairs, but also by linear processes, generating unwanted photonic noise. To characterize the single-photon spectrum and to identify the origins of noise, we program the WSS as a tunable band-pass filter operating from $191.289$ to $196.239 ~{\rm THz}$ with a step size of $33 ~{\rm GHz}$ and a bandwidth of $20 ~{\rm GHz}$.
	
	\begin{figure}[htb]
		\centering
		\includegraphics[width=0.45\textwidth]{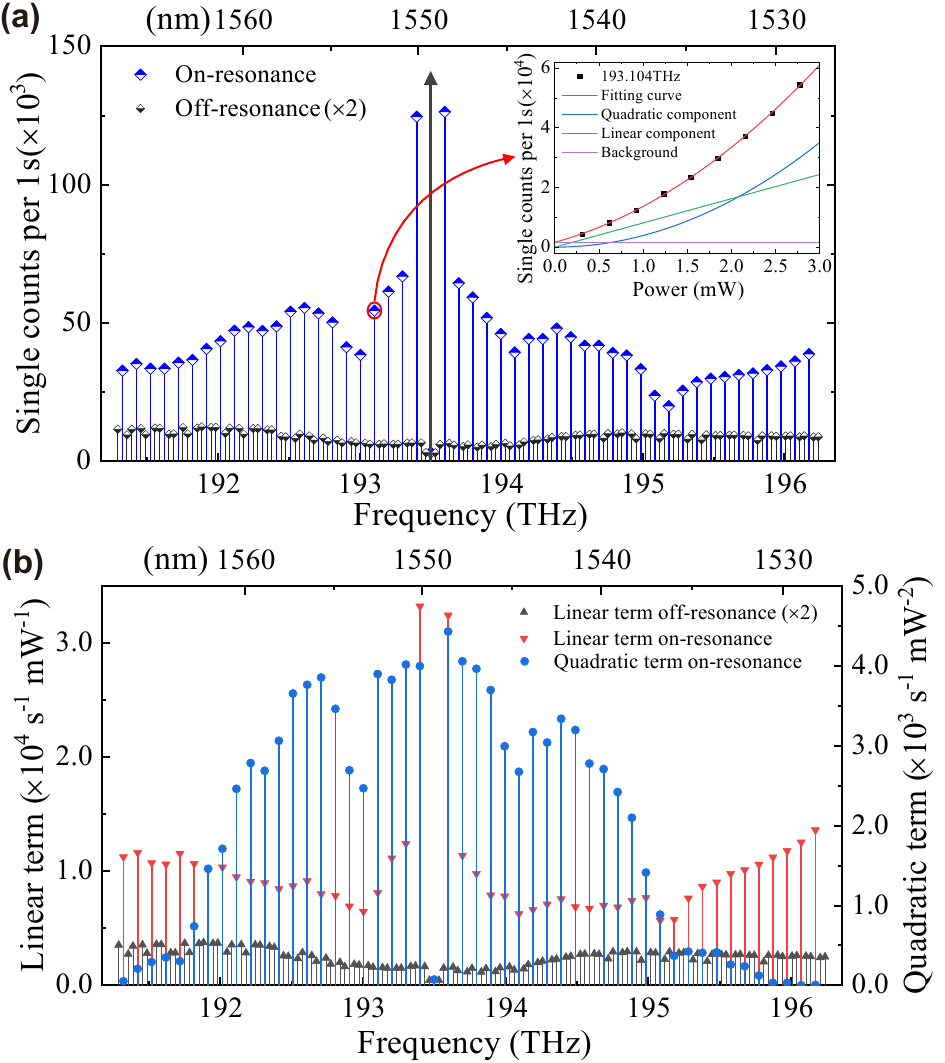}
		\caption{\label{Fig3}(a) Single-photon spectrum of the MRR for both on- and off-resonance cases. The minimum frequency separation is $33 ~{\rm GHz}$ for different channels and there are two off-resonance lines between every two adjacent resonance lines. The inset shows the single counts of a comb line (channel centered at $193.104 ~{\rm Hz}$) as a function of the pump power level. (b) The linear and quadratic terms can be separated and used to analyze photon noise by fitting each frequency line.  The values for the nonresonant cases are multiplied by 2 in (a) and (b).}
	\end{figure}
	
	FIG.\ref{Fig3}(a) demonstrates the measured single-photon spectrum, which reveals a clear difference in the single counts between the on-resonance and off-resonance cases. The inset shows the single counts as a function of pump power at a frequency channel of $193.104 ~{\rm Hz}$ (on resonance), and the data are fitted using the form $aP^2+bP+c$, where the quadratic contribution $aP^2$ comes from the photons generated by the SFWM, the linear contribution $bP$ comes from the photonic noise generated by the spontaneous Raman scattering \cite{Lin2007Photon,Reimer2016Generation}, and the constant term $c$ comes from the dark count (include the noise generated by WSS). We get $a=3.90 \times 10^3 ~{\rm s^{-1} mW^{-2}}$ for the quadratic term and $b=8.12 \times 10^3 ~{\rm s^{-1} mW^{-1}}$ for the linear term in this comb line. In comparison, the linear fit to a nearby frequency channel ($193.137 ~{\rm THz}$, off-resonance) shows an off-resonance linear term of $7.73 \times 10^2 ~{\rm s^{-1} mW^{-1}}$. In FIG.\ref{Fig3}(b), all frequency channels are analyzed and the linear and quadratic terms are extracted. For the $101$ off-resonance frequency channels, we obtain an average linear term coefficient of $1.22 \times 10^3  ~{\rm s^{-1} mW^{-1}}$, while for the on-resonance frequency channels (except the pump channel) it increases to $1.01 \times 10^4 ~{\rm s^{-1} mW^{-1}}$. That is, the photonic noise generated outside the ring is much smaller than that at on resonance, proving that there is additional photonic noise generated inside the MRR  \cite{Fan2023Multi, Samara2019amplified}. We also found that the average value of the linear term in the low-frequency part of the nonresonant channel is $1.32 \times 10^3  ~{\rm s^{-1} mW^{-1}}$, which is slightly higher than that in the high-frequency part with an average value of $1.14 \times 10^3  ~{\rm s^{-1} mW^{-1}}$, indicating the asymmetry of the noise spectrum, i.e., the Stokes part is larger than the anti-Stokes part \cite{Lin2007Photon}. For the quadratic term in the on-resonance case, we found that only $16$ pairs of frequency channels in a range of $3.168 ~{\rm THz}$ are highly efficient for generating correlated photon pairs, while the other channels are dominated by photonic noise, revealing a photon-pair generation bandwidth of approximately $ 25.4 ~{\rm nm}$ due to anomalous group-delay dispersion (GVD) at the excitation wavelength \cite{Zhang2019Generation}.
	
	\begin{figure}[htb]
		\centering
		\includegraphics[width=0.45\textwidth]{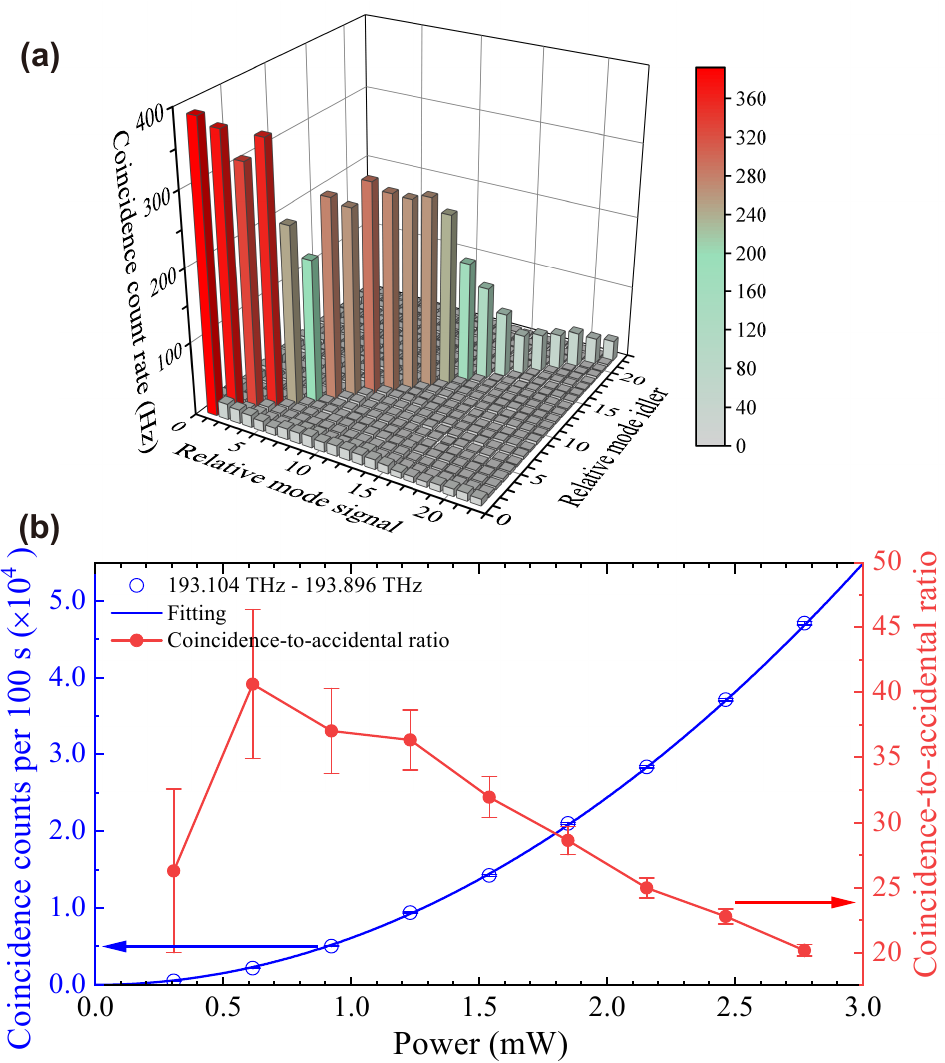}
		\caption{\label{Fig4}(a) Joint spectral intensity for $22$ accessible signal-idler pairs. (b) Coincidence counts and the calculated coincidence-to-accidental ratio (CAR) versus the pump power. Error bars are derived from Poissonian statistics.}
	\end{figure}
	
\begin{figure}[htb]
		\centering
		\includegraphics[width=0.45\textwidth]{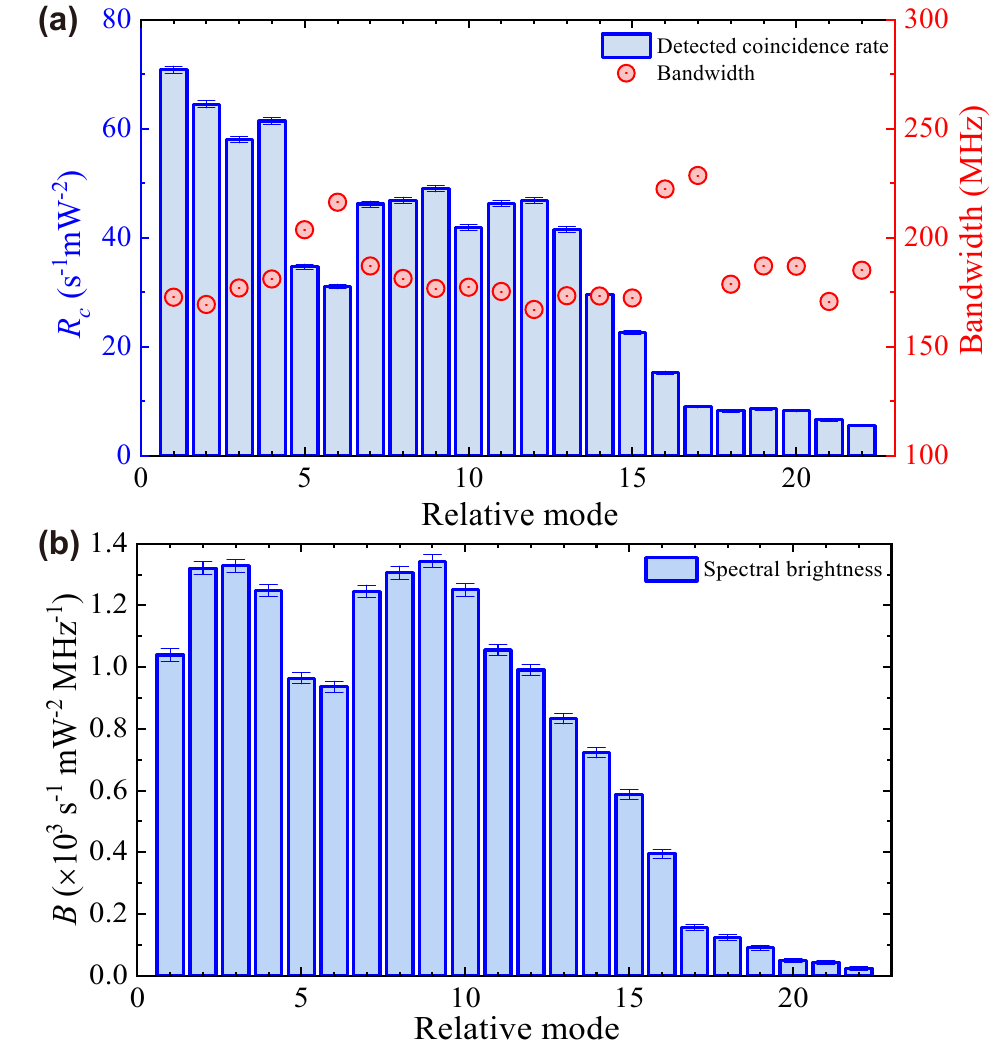}
		\caption{\label{Fig5}(a) Detected net coincidence rate $R_c$ and bandwidth for the respective frequency pairs. The bandwidth is derived from the temporal correlation histogram, and agrees with classical characterization results. (b) Spectral brightness $B$ for all $22$ pairs. Error bars are derived from Poissonian statistics.}
	\end{figure}

	Satisfying the energy conservation, the photon pairs generated by the SFWM process are spectrally symmetrically distributed on both sides of the pump, which is a multiwavelength two-photon entanglement source. To observe the frequency correlations over a wide spectral range, we characterized the joint spectral intensity (JSI) between combinations of frequency modes spanning a $22\times22$ Hilbert space \cite{Imany2018}. The JSI with channel bandwidth of $20 ~{\rm GHz}$ and channel spacing of $99 ~{\rm GHz}$ is shown in FIG.\ref{Fig4}(a). As expected, there is a strong correlation between the frequency-symmetric comb lines on the diagonal, reflecting the energy conservation during the SFWM process. We also noticed that the coincidence counts of the $17th$ to $22nd$ channel pairs are relatively low, which is consistent with the relative low quadratic term shown in FIG.\ref{Fig3}(b). For frequency-correlated photon pairs, the coincidence counts (CC) and the coincidence-to-accidental ratio (CAR) are measured. Taking a particular frequency pair ($193.104 - 193.896 ~{\rm THz}$) shown in FIG.\ref{Fig4}(b) as an example, the maximum CAR achievable is $41$, which can be improved by reducing signal and idler photon losses. For the $22$ pairs of correlated channels on the diagonal, we measured the coincidence counts at different pump-power levels, and derived the detected net coincidence rate $R_c$ with the unit of $~{\rm s^{-1} mW^{-2}}$. The results are shown in FIG.\ref{Fig5}(a). Following the method described in Ref. \cite{Wen2022}, we obtained an average bandwidth of $184.58 ~{\rm MHz}$ for each pair, which is in good agreement with the linewidth of the MRR. FIG.\ref{Fig5}(b) shows the spectral brightness $B$ with a unit of $~{\rm s^{-1} mW^{-2} MHz^{-1}}$, which is independent of collection efficiency. Refer to Appendix E for the detailed definition of the detected net coincidence rate $R_c$ and the spectral brightness $B$.
	
\begin{figure}[htb]
		\centering
		\includegraphics[width=0.4\textwidth]{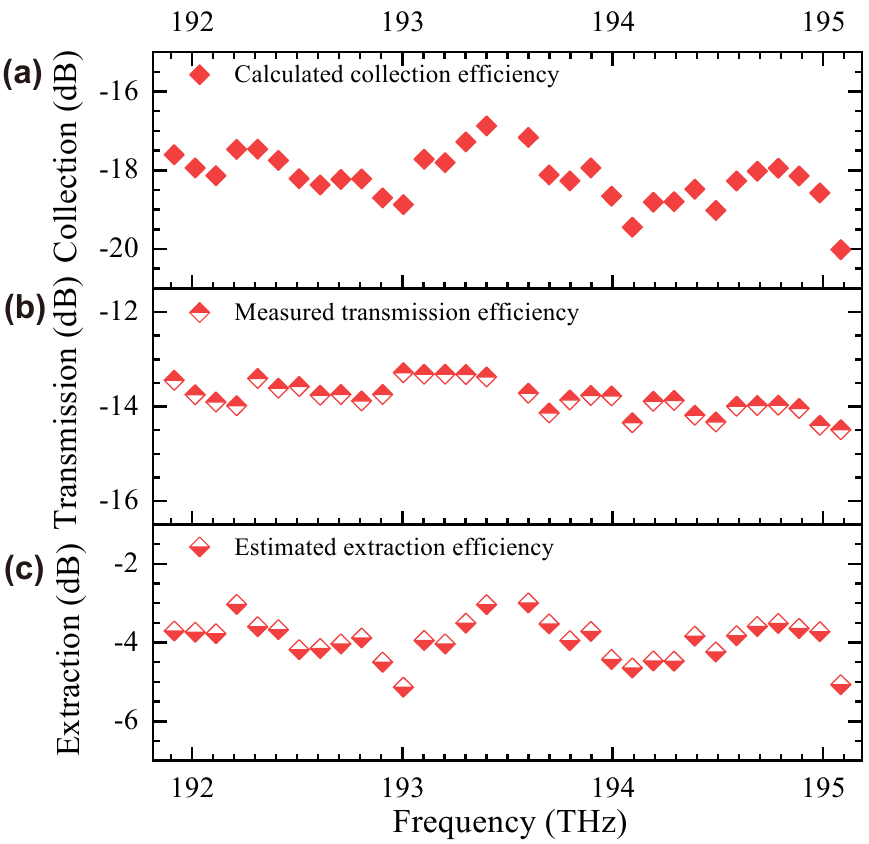}
		\caption{\label{Fig6}(a) The calculated collection efficiency, (b) the measured transmission efficiency (include the fiber-to-chip coupling efficiency), and (c) the estimated ring-to-bus extraction efficiency for different frequency channels.}
	\end{figure}

    The collection efficiency \cite{Engin2013photon,Li2023discrete} of the signal (idler) channel shown in FIG.\ref{Fig6}(a) can be obtained by dividing the biphoton coincidence count rate $R_c$ by the single-photon count rate $R_i (R_s)$ of the idler (signal) channel, which contains the ring-to-bus extraction efficiency, the total transmission efficiency [include the fiber-to-chip coupling efficiency (approximately $50\%$)] and the detection efficiency (approximately $90\%$). For the 16 generated frequency pairs, we measured the transmission losses of the signal and idler channels, as shown in FIG.\ref{Fig6}(b). Subtracting the total transmission loss and detection loss from the collection efficiency gives us the ring-to-bus waveguide extraction efficiency shown in FIG.\ref{Fig6}(c), with an average value of $40.79\%$ ($-3.92 ~{\rm dB}$).Further optimization of the coupling conditions to slightly overcoupling can improve the extraction efficiency and also the photon-pair generation rate \cite{Fan2023Multi}.

	The source brightness decays rapidly due to the limited SFWM bandwidth, which is consistent with the results shown in FIG.\ref{Fig3}(b). And the single counts becoming increasingly dominated by noise as the frequency pair moves away from the pump. Meanwhile, the first frequency channel pair, which is closest to the pump, is affected by the noise photons, reducing its quality. Although the observed fidelity is greater than $81\%$ for all channel pairs, we recognize that further research is needed to increase the number and quality of channel pairs. For the SFWM bandwidth, the GVD in waveguide can be tailored to achieve anomalous GVD over a broader bandwidth \cite{Turner2006Tailored}. The effect of higher pump power level on the bandwidth also needs to be investigated. Increasing the number of channel pairs can also be achieved by increasing the cavity length of the microresonator, i.e., reducing the FSR, for a fixed bandwidth. With advances in integrated photonics, the entire platform can be integrated on a monolithic chip, reducing losses. Cryogenic cooling of the chip can suppress the photon noise generated in the chip \cite{Li2005optical, Feng2023Entanglement}. All the above-mentioned points could improve the signal-to-noise ratio, and enables higher-quality quantum light sources.
	
	\section{Discussion and Conclusion}
	
	In contrast to the generation of photon pairs in nonlinear bulk crystals or waveguides with subsequent narrow-band spectral filtering, microresonators with multiple frequency nature can efficiently generate discrete, equally spaced frequency combs. The extensibility of the frequency modes of QFC, with the additional entanglement of other DOF, allows the generation of frequency-multiplexed heralded photons and multiphoton, high-dimensional and hyperentangled states \cite{Kues2017chip, Reimer2019high}. Meanwhile, the polarization entanglement source uses the well-established PAM for entanglement analysis, eliminating the requirement for the phase stability of the interferometer in the energy-time entanglement analyzers. And the use of higher $Q$-factor microresonators can lead to narrower photon bandwidth and higher photon-pair generation rates, resulting in higher spectral brightness, which may bring quantum light sources into real-world application scenarios.
	
	The rapid development of integrated photonics \cite{wang2020integrated, Wang2021Integrated, Pelucchi2022potential} in recent years has made it possible to realize higher $Q$-factor microresonators. Meanwhile, integrated photonic devices, such as polarization beam splitter and polarization rotator \cite{Sacher2014Polarization, Hou2019On}, have been rapidly developed, which provides a promising solution for fully integrated, compact, stable, and scalable polarization entanglement sources. We also noticed that the SFWM bandwidth of our MRR is approximately $3 ~{\rm THz}$, which may attribute to the relatively low pump power level and the nonoptimal dispersion engineering of the waveguide. Cryogenic cooling, better dispersion engineering, smaller FSR (narrower frequency-mode spacing), can provide more frequency modes with efficient pair generation rate, which is very helpful for realizing large-scale QNs based on WDM.
	
	In summary, we have demonstrated a polarization-entangled biphoton QFC by placing an integrated silicon-nitride MRR within a Sagnac interferometer. We show that our MRR-based Sagnac-configured source can simultaneously generate $22$ parallel frequency pairs in a range of approximately $4 ~{\rm THz}$. Using quantum state tomography to quantify the entanglement of each channel pair, the observed fidelity exceeds $81\%$ in all frequency ranges. The freedom of the Sagnac interferometer from phase control between different paths, together with the feedback control we developed based on hysteretic behavior, allows our source to be highly stable against external perturbations. In addition, the presented setup has the potential to be integrated into a single chip, which can further improve the stability of the system. Our research provides a scalable and integrated platform for the generation of various entangled photonic states, as well as for the experimental realization of high-dimensional quantum computing and large-scale QNs.
	
	\begin{acknowledgments}
		This research was supported by the National Key Research and Development Program of China (Grants No. 2022YFE0137000, No. 2019YFA0308704), the Leading-Edge Technology Program of Jiangsu Natural Science Foundation (Grant No. BK20192001), the Fundamental Research Funds for the Central Universities, the Innovation Program for Quantum Science and Technology (Grants No. 2021ZD0300700 and No. 2021ZD0301500), and the program B for Outstanding PhD candidate of Nanjing University.
	\end{acknowledgments}
	
	\appendix
	
	\renewcommand\thefigure{\Alph{section}\arabic{figure}}
	\setcounter{figure}{0}
	
	\section{Cavity resonances}
	
	\begin{figure}[htb]
		\centering
		\includegraphics[width=0.45\textwidth]{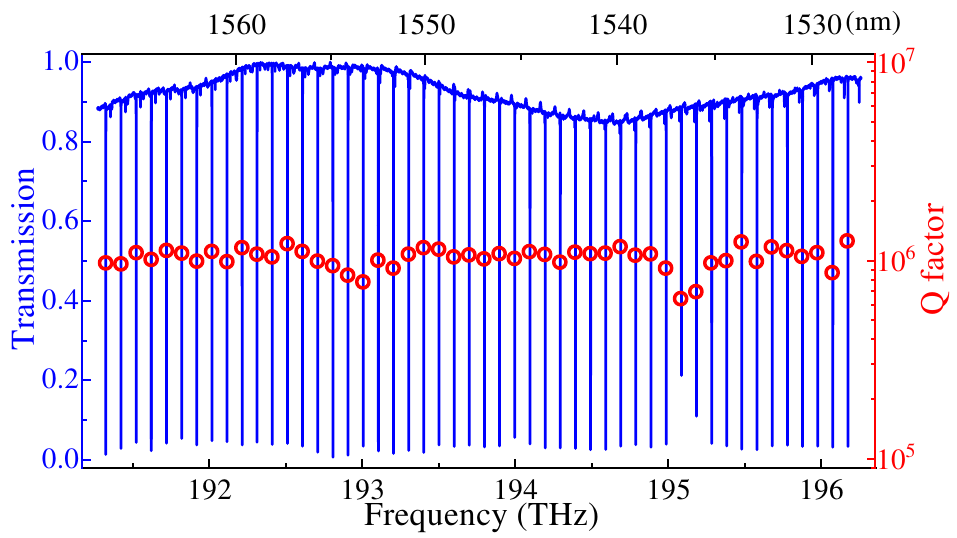}
		\caption{\label{FigS1}Transmission spectrum and $Q$ factors for $50$ resonances from $191.269$ to $196.259 ~{\rm THz}$, with a free spectral range of approximately $99 ~{\rm GHz}$.}
	\end{figure}
	
	The transmission spectrum of the MRR around the telecom C-band is measured using a tunable laser (Santec TSL-570) under near-cold cavity conditions, and a total of $50$ resonances were obtained from $191.269$ to $196.259 ~{\rm THz}$, as shown in FIG.\ref{FigS1}. We derived quality factors for all measured resonances, which exhibited an average FWHM of $190.41 ~{\rm MHz}$ and an average $Q$ factor of $1.03\times 10^{6}$. An experimental value for the free spectral range (FSR) of $99.03 ~{\rm GHz}$ is obtained by averaging the frequency spacings of adjacent resonances.
	
	\section{Thermal control of the resonances}
	
	\begin{figure}[htb]
		\centering
		\includegraphics[width=0.45\textwidth]{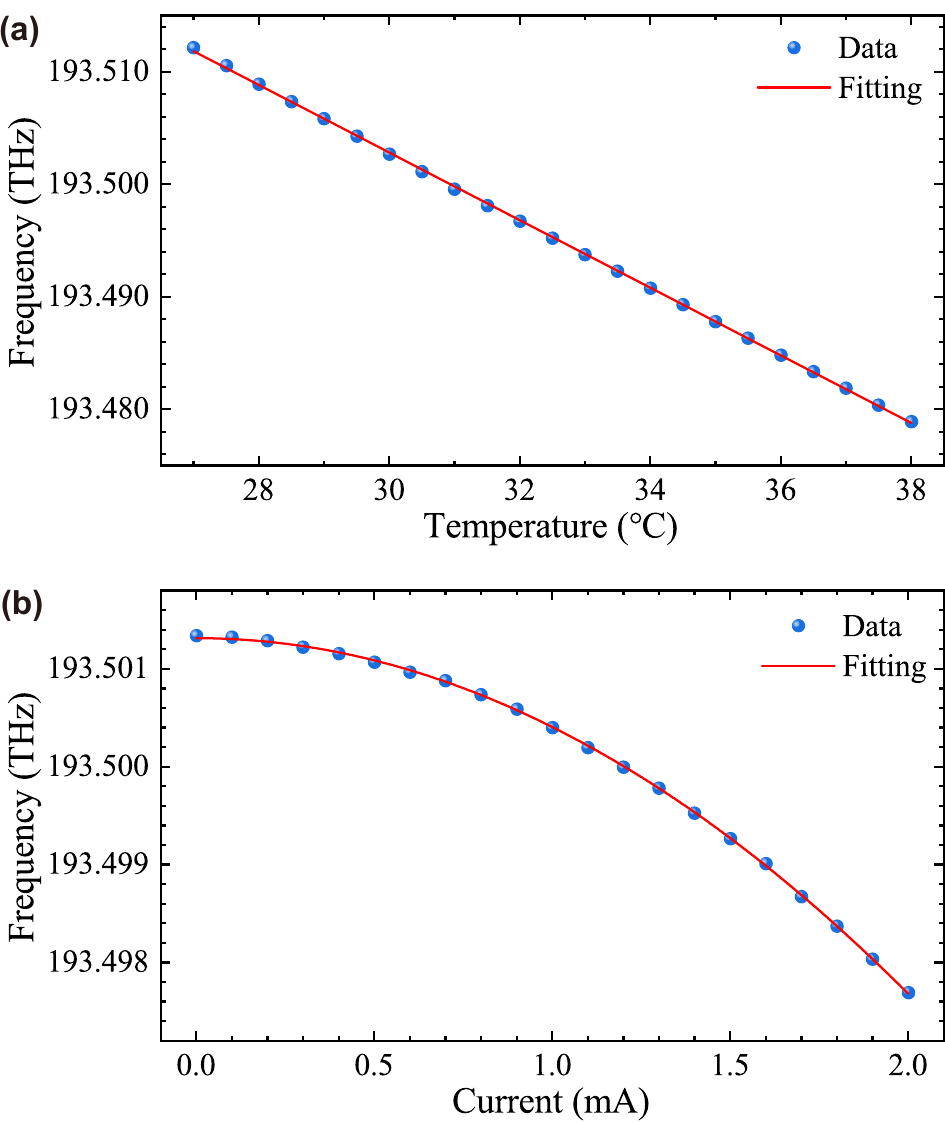}
		\caption{\label{FigS2}Resonant frequency of the MRR as a function of (a) chip temperature, (b) current applied to the integrated microheater.}
	\end{figure}
	
	For the thermal control of the resonance of the MRR, we have designed an integrated microheater placed at the top of the ring waveguide, which has an electrical resistance of about $2050 ~{\rm \Omega}$. Due to the thermo-optical effect, electrically driving the integrated heater changes the refractive index of the waveguide, thereby adjusting the spectral positions of the cavity resonances. The chip is mounted on a copper plate with a thermoelectric cooler (TEC) and a negative temperature coefficient (NTC) thermistor connected to a high-performance digital temperature controller (Thorlabs TED4015) to stabilize the temperature ($<0.001~^{\circ} \mathrm{C}$) of the entire chip. When the pump power is relatively low (i.e. near a cold cavity), the cavity resonance exhibits a Lorentzian lineshape. And we find a resonance shift of about $-3.01 ~{\rm GHz / ^{\circ} \mathrm{C}}$ for the temperature tuning and about $-0.91 ~{\rm GHz/mA^2}$ or $-0.44 ~{\rm GHz/mW}$ for the current tuning, according to FIG.\ref{FigS2}. 
	
	\section{The hysteretic behavior}
	
\begin{figure}[H]
		\centering
		\includegraphics[width=0.45\textwidth]{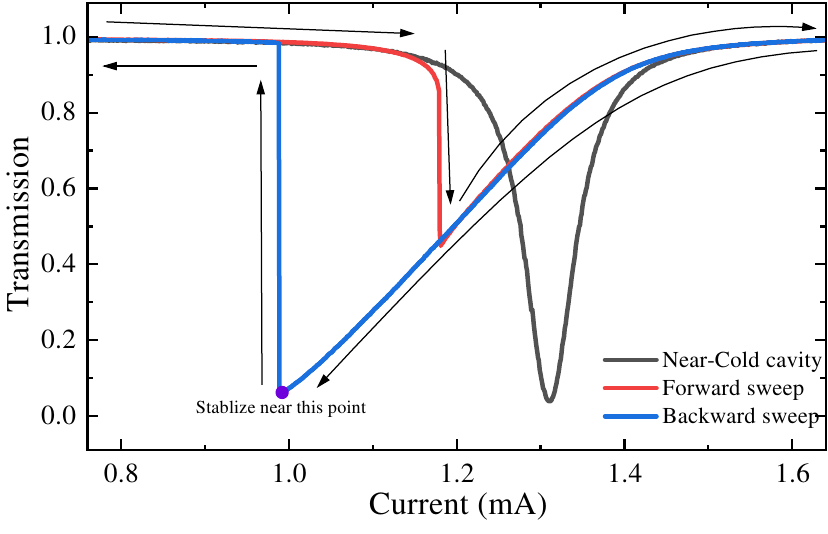}
		\caption{\label{FigS3}Hysteresis behavior of the pump resonance with the laser frequency fixed at $193.5 ~{\rm THz}$. The cavity resonance exhibits a Lorentzian lineshape under near-cold cavity condition ($0.015 ~{\rm mW}$). The current of the microheater is tuned forward and backward with a pump power of $2.773 ~{\rm mW}$.}
	\end{figure}

	FIG.\ref{FigS3} shows the hysteresis behavior for different pump levels, where the pump frequency is fixed at $193.5 ~{\rm THz}$ and the microheater current is tuned so that the cavity resonance sweeps across the pump line. Note that the cavity Lorentz is initially to the left of the pump line. A forward current sweep (increase in current) is equivalent to a backward wavelength sweep (decrease in pump wavelength) and vice versa. As the current decreases (corresponding to the pump wavelength approaching the cavity resonance from the left), the heat flowing into the cavity increases and the cavity begins to heat up, drifting the cavity resonance away from the cold-cavity resonance wavelength. The rise of cavity temperature stops when the absorption is maximal and cannot be further compensated by the heater dissipation. The cavity resonance cannot be pushed any further and the pump power in the cavity drops quickly. However, as the current increases, the pump approaches the cavity resonance from the right side, and the upward shift of the cavity resonance is accelerated as the cavity heats up. The scan rate across the cavity lineshape is sufficiently fast and prevents the resonator from fully “charging”, resulting in the measured transmission not reaching the minimum.
	
	\section{Long-term stable operation of the source}

\begin{figure}[H]
		\centering
		\includegraphics[width=0.45\textwidth]{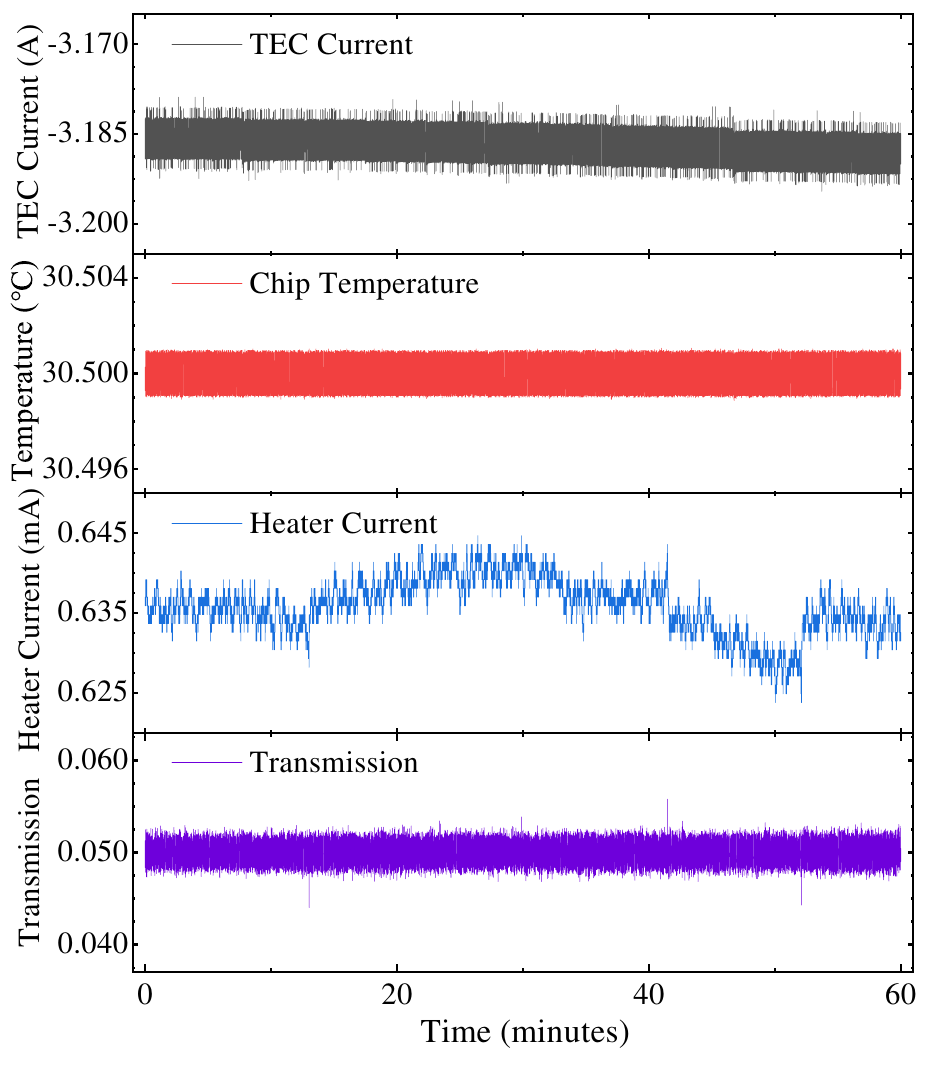}
		\caption{\label{FigS4}Resonance stability with the help of chip temperature control and microheater-based feedback control.}
	\end{figure}

	Stable pumping of a high-$Q$ micro-resonator with narrow linewidth needs to overcome environmental perturbations caused by temperature, wavelength, power, etc. And the existence of hysteresis behavior makes it easier for the micro-resonator to drop out of resonance. Whereas, the broadening of the cavity lineshape provides a larger range to stabilize the cavity near the resonance point (i.e., the minimum transmission point) of the pump. We monitor the pump power after the postfilter, then decrease the current to push the cavity resonance toward the pump line and stay just before the minimum transmission point. If there is a small decrease in pump transmission, it means that the heat flowing into the cavity is increasing, we can increase the current to push the cavity resonance away from the pump to cool the cavity. Similarly, the current should decrease when an increase in transmittance is detected. As shown in FIG.\ref{FigS4}, this feedback control of the cavity resonances, combined with the temperature control of the chip, allows us to stabilize the cavity near the resonance with a transmission of $0.05$, corresponding to an extinction ratio of approximately $13 ~{\rm dB}$.
	
	\section{Spectral brightness and collection efficiency}
	In our experiment, the number of signal and idler photons detected by SNSPDs per second ($ N_s $ and $ N_i $) are not only generated by the SFWM process, but also by the spontaneous Raman scattering \cite{Lin2007Photon} and the dark counts ($ N_{\mathrm{DC},s} $ and $ N_{\mathrm{DC},i} $), and the subscripts $ s $ and $ i $ represent signal and idler, respectively. Taking into account the intrinsic photon-pair generation rate ($ R_{\mathrm{PGR}} $) and the collection efficiency of each channel ($ \eta_s $ and $ \eta_i $), these parameters are related to each other through the following formulas \cite{Tanzilli2001highly, Engin2013photon}:
	\begin{equation}
		\begin{split}
			N_s &=\eta _s\times R_{\mathrm{PGR}}\times P^2+R_{\mathrm{RS},s}\times P+N_{\mathrm{DC},s}\\
			&=R_s\times P^2+R_{\mathrm{RS},s}\times P+N_{\mathrm{DC},s},\\
			N_i &=\eta _i\times R_{\mathrm{PGR}}\times P^2+R_{\mathrm{RS},i}\times P+N_{\mathrm{DC},i}\\
			&=R_i\times P^2+R_{\mathrm{RS},i}\times P+N_{\mathrm{DC},i},
		\end{split}
	\end{equation}
	where $P$ is the input power of pump, the collection efficiencies $ \eta_s $ and $ \eta_i $ contains the ring-to-bus waveguide extraction efficiency, the total transmission efficiency [include the fiber-to-chip coupling efficiency (approximately $50\%$)] and the detection efficiency (approximately $90\%$), the quadratic coefficient $ R_{s(i)} = \eta_{s(i)} \times R_{\mathrm{PGR}} $ here is the detected single-photon count rate associated with SFWM in the unit of $ s^{-1} mW^{-2} $. The linear coefficient $ R_{\mathrm{RS},s(i)} $ is the noise term that associated with Raman scattering, which includes the contributions of the input and output fiber, the silicon-nitride bus waveguide and the microring. For the nonresonant case, the quadratic term is close to zero, and there is almost no contribution of the microring to the noise in the linear coefficient.	Meanwhile, the detected coincidence counts per second of the photon pairs satisfy
	\begin{equation}
		\begin{split}
			N_c &=\eta _s\times \eta _i\times R_{\mathrm{PGR}}\times P^2+N_{\mathrm{AC}}\\
			&=R_c\times P^2+N_{\mathrm{AC}},
		\end{split}
	\end{equation}
	where $N_{\mathrm{AC}}$ is the accidental coincidence count rate in the coincidence window, and the quadratic coefficient $ R_c=\eta _s \times \eta _i \times R_{\mathrm{PGR}} $ is the detected net coincidence rate in the unit of $ s^{-1} mW^{-2} $, which includes the losses experienced by signal and idler photons. Thus, the intrinsic pair generation rate, which excludes all losses, can be calculated by the following formula:
	\begin{equation}
		R_{\mathrm{PGR}} = \frac{R_s \times R_i}{R_c}\left( s^{-1}mW^{-2} \right).
	\end{equation}
	Spectral brightness $B$ is defined as the number of photon pairs generated by a light source per second per unit pump power per megahertz of bandwidth, and can be calculated by dividing the PGR by the corresponding bandwidth with a unit of $ s^{-1} mW^{-2} MHz^{-1}$. The loss on the signal and idler photons, that is, the collection efficiency, can be derived by:
	\begin{equation}
		\eta _s = \frac{R_c}{R_i} \ and \ \eta _i = \frac{R_c}{R_s}.
	\end{equation}
	The ring-to-bus waveguide coupling loss can then be estimated by subtracting the experimentally measured total transmission loss and detection loss from the collection loss.

    \section{The sandwiched QWP-HWP-QWP setup}
    In the sandwiched "QWP-HWP-QWP" waveplate combination, the angle of the two QHP is fixed at $45^{\circ}$, and the angle of the middle half-wave plate is marked as $\varphi$. Then the Jones matrix after the input state $\left[ 1;e^{i\theta} \right]$ passes through "QWP-HWP-QWP" can be expressed as:
    \begin{equation}
    \begin{split}
        \mathbf{J}_{\mathrm{output}}
        &= \mathbf{J}_{\mathrm{QWP}}(\frac{\pi}{4})\times \mathbf{J}_{\mathrm{HWP}}(\phi )\times \mathbf{J}_{\mathrm{QWP}}(\frac{\pi}{4})\times
        \begin{pmatrix}
            1\\
            \mathrm{e}^{\mathrm{i}\theta}
        \end{pmatrix}\\
        &= \begin{pmatrix}
            \mathrm{e}^{\mathrm{i}\left( \frac{\pi}{2}-2\phi \right)}&		0\\
	0&		\mathrm{e}^{\mathrm{i}\left( 2\phi -\frac{\pi}{2} \right)}\\
        \end{pmatrix}
        \begin{pmatrix}
            1\\
            \mathrm{e}^{\mathrm{i}\theta}
        \end{pmatrix}\\
        &= \begin{pmatrix}
            \mathrm{e}^{\mathrm{i}\left( \frac{\pi}{2}-2\phi \right)}\\
	\mathrm{e}^{\mathrm{i}\left( 2\phi -\frac{\pi}{2}+\theta \right)}
        \end{pmatrix}\\
        &= \mathrm{e}^{\mathrm{i}\left( \frac{\pi}{2}-2\phi \right)}
        \begin{pmatrix}
            1\\
            \mathrm{e}^{\mathrm{i}\left( 4\phi +\theta -\pi \right)}
        \end{pmatrix}.
    \end{split}
    \end{equation}
    In order to compensate for the relative phase difference $\theta$ between $\left| HH \right>$ and $\left| VV \right>$, the phase $\theta$ and the angle of the HWP $\varphi$ need to satisfy:
    \begin{equation}
        4\phi +\theta =\pi.
    \end{equation}
    That is, the phase difference $\theta$ between $\left| HH \right>$ and $\left| VV \right>$ to be compensated is four times the angle of the HWP $\varphi$ (plus an additional constant).
    	
	\bibliography{SiN-Sagnac}
	
\end{document}